\documentclass[12pt]{article}
\usepackage{epsfig,latexsym,amssymb}

\newcommand{\ket}[1]{|{#1}\rangle}
\newcommand{\ii}{\mathrm{i}}
\newcommand{\ee}{\mathrm{e}}
\newcommand{\eq}[1]{Eq.~(\ref{#1})}
\newcommand{\secn}[1]{Section~\ref{#1}}
\def\be{\begin{equation}}
\def\beq{\begin{equation}}
\def\eeq{\end{equation}}
\newcommand{\en}{\end{equation}}
\def\ba{\begin{eqnarray}}
\def\bea{\begin{eqnarray}}
\def\ea{\end{eqnarray}}
\def\eea{\end{eqnarray}}
\newcommand{\eqa}{\begin{eqnarray}}
\newcommand{\ena}{\end{eqnarray}}

\def\Nc{{\cal{N}}}
\def\Ac{{\cal{A}}}
\def\ch{{\rm ch}}
\def\Ch{{\rm Ch}}
\def\CH{{\rm CH}}
\def\tr{{\rm tr}}

\def\n{\mathbf{n}}
\def\r{\mathbf{r}}

\begin{document}
\begin{titlepage}
\vskip0.5cm
\begin{flushright}
DFTT 6/01\\
\end{flushright}
\vskip 2.4cm
\begin{center}
{\Large\bf
Branched Coverings and Interacting  Matrix Strings in Two Dimensions}
\end{center}
\vskip 1.cm
\centerline{
M. Bill\'o$^a$,
A. D'Adda$^a$ and P. Provero$^{b,a}$}
\vskip0.6cm
\centerline{\sl  $^a$ Dipartimento di Fisica
Teorica dell'Universit\`a di Torino and}
\centerline{\sl Istituto Nazionale di Fisica Nucleare, Sezione di Torino}
\centerline{\sl via P.Giuria 1, I-10125 Torino, Italy}
\vskip 0.2cm
\centerline{\sl $^{b}$ Dipartimento di Scienze e Tecnologie Avanzate}
\centerline{\sl Universit\`a del Piemonte Orientale}
\centerline{\sl I-15100 Alessandria, Italy
\footnote{e--mail: {\tt billo, dadda, provero@to.infn.it}}}
\vskip 2cm
\begin{abstract}
We construct the lattice gauge theory of the group $\cal{G}_N$, the semidirect
product of the permutation group $S_N$ with  U$(1)^N$, on an arbitrary Riemann
surface. This theory describes the branched coverings of a two-dimensional 
target surface by strings carrying a U$(1)$ gauge field on the world sheet.
These are the non-supersymmetric Matrix Strings that arise  in the unitary
gauge quantization of a generalized two-dimensional Yang-Mills theory. By
classifying the irreducible representations of $\cal{G}_N$, we  give the most
general formulation of the lattice gauge theory of $\cal{G}_N$, which includes
arbitrary branching points on the world sheet and describes the splitting and
joining of strings. 
\end{abstract}
\end{titlepage}
\setcounter{footnote}{0}
\def\thefootnote{\arabic{footnote}}

\section{Introduction}
The relation between two dimensional Yang-Mills theories in the large $N_c$
limit and two dimensional string theories was established  by Gross and Taylor
in a series of papers
~\cite{
Gross:1993tu,
Gross:1993hu,Gross:1993yt}. 
They showed  that the coefficients of the $\frac{1}{N_c}$ expansion of the YM
partition function, with gauge group $SU(N_c)$ on a Riemann surface${\cal M}$
of area $\Ac$ and  arbitrary genus $G$ count the number of string 
configurations without folds, namely the number of branched coverings of the
surface. As a matter of fact two dimensional YM does not give exactly a pure
theory of coverings  for two reasons: the presence of two distinct chiral
sectors,which are weakly coupled by pointlike tubes and correspond to the two
possible orientations of the world sheet, and the presence, for $G>1$, of the
so called $\Omega^{-1}$ points, whose geometric meaning was eventually
clarified in ~\cite{Cordes:1997sd,Cordes:1995fc}.
 \par
A lattice gauge theory of complex $N_c \times N_c$ matrices that describes, in
the large $N_c$ limit, a pure theory of branched coverings with just one chiral
sector and without $\Omega^{-1}$ points was later discovered in
~\cite{Kostov:1997bs,Kostov:1998bn}.  In fact, when expanded in powers of
$q=e^{-\Ac}$, the partition function of this complex matrix model can be
regarded as the generating function of the number of branched coverings
$Z_{G,N}$ that wrap around $\cal M$ $N$ times. Actually for finite $N_c$ only
the first $N_c$ coefficient of the expansion correctly reproduce the
corresponding $Z_{G,N}$ and the generating function is only reproduced in the
large $N_c$ limit.  A similar property holds for two dimensional YM theories
~\cite{Baez:1994gk}.
\par
For fixed $N$, the number of branched coverings $Z_{G,N}$ is itself the
partition function of a lattice gauge theory
\cite{Gross:1993hu,Kostov:1997bs,Kostov:1998bn} whose gauge group is the
symmetric group $S_N$ and whose plaquette action is the standard heat-kernel
action for $S_N$:
\begin{equation}
\ee^{-S_{\rm pl}(P,{\cal A}_{\rm pl})} = \frac{1}{(N!)^2}\sum_r d_r \ch_r(P)
\ee^{{\cal A}_{\rm pl}\,g_r}~,
\label{heatker}
\end{equation}
where $P \in S_N$ and $r$ labels the representations of $S_N$ of characters
$\ch_r$ and dimensions $d_r$; ${\cal A}_{\rm pl}$ is the area of the plaquette
and the arbitrary coefficients $g_r$ carry all the information about the
density of all different types of branch points. The relation between the
lattice matrix theory of ~\cite{Kostov:1998bn} and the $S_N$ gauge theory is a
direct consequence of Frobenius formula, as shown in
~\cite{Kostov:1997bs,Kostov:1998bn}.
\par
A completely different way of obtainig a string theory from generalized two
dimensional YM theories was proposed in ~\cite{Billo:1999fb,Billo:2000ts}. The
mechanism is analogue to the one through which string configurations emerge in
"Matrix string theory"~\cite{Dijkgraaf:1997vv} and it is based on the
quantization of a generalized\footnote{Generalized YM theories have been
discussed in ~\cite{Douglas:1994pq,Ganor:1995bq}.} YM theory in a unitary
gauge. More precisely, one starts from the generalized action
\begin{equation}
S_{\rm gen} = \int_{\cal M} d\mu\, V(B)-
\ii\, \tr \int_{\cal M} (d A -\ii A\wedge A~) B~,
\label{genact}
\end{equation}
where the gauge group is U$(N)$ and $V(B)$ is an arbitrary gauge invariant
potential of the $N \times N$ hermitian matrix%
\footnote{We denote here the number of colours  by $N$ as in this case it
coincides with the number of times the world sheet of the string wraps round
the target space. This marks an important difference between this approach and
the one of \cite{Gross:1993hu} or \cite{Kostov:1997bs}}
$B$. One chooses the unitary gauge in which the matrix $B$ is diagonal. This
gauge choice is incomplete, as it leaves a residual U$(1)^N$ gauge invariance.
It is also affected by Gribov ambiguities as at each point of the space-time
manifold the gauge is determined up to an arbitrary permutation of the
eigenvalues of $B$. A global smooth diagonalization is then in general not
possible because, as one goes round a loop non contractible to a point, the
eigenvalues of $B$ may be subjected to a permutation. This gives origin to
twisted sectors, which are in one-to-one correspondence with the homomorphisms
$\Pi_1(x|{\cal M}) \rightarrow S_N$ of the  homotopy group of ${\cal M}$ onto
the symmetric group $S_N$, namely with the unbranched $N$-coverings of ${\cal
M}$.
\par
In four space-time dimensions unitary gauges lead to divergences when two
eigenvalues coincide, and the theory is apparently non-renormalizable
~\cite{'tHooft:1974hx} in these  gauges. In two dimensions these divergences
still occur, but they are exactly cancelled in flat space-time. In fact there
is an exact cancellation, due to a fermionic symmetry, between the
contributions of the non-diagonal components of the gauge field and the
ghost-antighost system ~\cite{Billo:1999fb}. This fermionic symmetry is however
anomalous in presence of space-time curvature, and it appears that the only
consistent way to eliminate the resulting divergences is to make use of the
arbitrariness of the potential $V(B)$ in (\ref{genact}) and introduce in it a
term, which depends on the curvature, that cancels exactly the anomaly. With
this prescription the theory becomes completely abelian, consisting of $N$
U$(1)$ gauge theories, whose field strengths are the eigenvalues of $B$, and
which are only coupled, in each twisted sector, by the boundary conditions
induced by the homomorphism $\Pi_1(x|{\cal M}) \rightarrow S_N$. The original
generalized YM theory, with U$(N)$ gauge group, is then described by a string
theory (unbranched coverings) that covers $\cal M$ $N$ times, and has a U$(1)$
gauge theory on its world sheet.
\par
As shown in ~\cite{Billo:2000ts}, this theory can be formulated as a lattice
gauge theory whose gauge group ${\cal G}_N$ is the semi-direct product of $S_N$
and U$(1)^N$ defined by the multiplication rule
\begin{equation}
(P,\varphi)\;(Q,\theta)=(PQ, \varphi+P\theta)~,
\label{multrule}
\end{equation}
where $(P,\varphi)$ can be represented as the $N \times N$ matrix
\begin{equation}
(P,\varphi)_{ij} = e^{\ii\varphi_i} \delta_{iP(j)}
\label{matrel}
\end{equation}
and $P$ is an element of $S_N$.
The twisted sectors considered in ~\cite{Billo:2000ts} did not include the
possibility of branch points. Correspondingly, the plaquette action in the
${\cal G}_N$ lattice gauge theory had the form
\begin{equation}
e^{-S_{\rm pl}} = \delta(P_{\rm pl})\sum_{n_i} \exp\left\{\sum_{i=1}^{N}
\left(\ii n_i \varphi^{(pl)}_i-{\cal A}_{\rm pl} v(n_i)\right)\right\}~,
\label{plact}
\end{equation}
where
$v(n_i)$ is the remaining part of $V(B)$ after the anomaly cancellation.
The plaquette action for
the $S_N$ subgroup in (\ref{plact})  is just a delta function $\delta(P_{pl})$;
this denotes the absence of branch points.
\par
The problem of introducing twisted sectors that allow branch points, so that
the strings carrying the U$(1)$ field can join and split, will be addressed in
this paper. From the point of view of the lattice gauge theory of  ${\cal G_N}$
this is equivalent to writing the most general heat kernel action for ${\cal
G_N}$ (as it is done in eq. (\ref{heatker}) for the group $S_N$). This entails
finding an explicit expression for the characters of the irreducible
representions of ${\cal G_N}$ in terms of the characters of $S_N$ and
U$(1)^N$.  From the point of view of the (generalized)  Yang-Mills theory, the
string interactions should originate from non-perturbative effects, in analogy
to what happens in the supersymmetric case of $\mathcal{N}=8$ super Yang--Mills
theory. In this case the string theory that arises in the diagonal
gauge%
\footnote{ 
Here the field which are diagonalized in the Yang-Mills strong
coupling limit are the eight scalar superpartners of the gauge field.}  
is the Matrix String of \cite{Dijkgraaf:1997vv}, which is believed to be
equivalent to the Type IIB string theory in the light-cone frame. The string
interactions \cite{Wynter:1997yb,Giddings:1999yd,Wynter:2000gj,Brax:2000zc,
Grignani:2000zm}
arise precisely because there are classical instantonic
configurations \cite{Bonelli:1998yt,Bonelli:1999wx,Bonelli:1999qa} which lead
to a structure of eigenvalues  describing a branched covering of the target
spacetime manifold. The string coupling constant $g_s$ is, consistently with
this picture,  proportional to $1/g_{\mathrm{YM}}^2$.
\par
In our case it is not yet clear how to explicitely derive analogous instantonic
configurations for the generalized Yang-Mills theory; nevertheless we can
investigate  the interacting theory, i.e. the theory of branched coverings,
irrespectively of its dynamical origin.
\section{$N$-coverings as $S_N$ gauge theory}
\label{sec:dualcov}
In this Section we will show that the partition function of
$N$-coverings of a general Riemann surface can be expressed as the
partition function of a lattice gauge theory, defined on a cell
decomposition of the surface, with the symmetric group
$S_N$ as the gauge group.
\par
A covering of a {\it target}  Riemann surface $\Sigma_T$ is essentially a
smooth map
\be
f:\Sigma_W\rightarrow \Sigma_T
\eeq
from a {\it covering} surface $\Sigma_W$ (the ``world-sheet'') to $\Sigma_T$
(the ``target'') such that each point of $\Sigma_T$ has exactly $N$
counterimages in $\Sigma_W$. Each of these counterimages is said to belong to a
different {\it sheet} of the covering surface. Consider now a path $\gamma$ in 
$\Sigma_T$, from a point $P_1$ to a point $P_2$. It is natural to define the
$N$ {\it liftings} $\tilde{\gamma}_i$ of $\gamma$ as the paths on $\Sigma_W$
such that
\be
f\circ \tilde{\gamma}_i=\gamma~.
\eeq
If the liftings of all closed, homotopically trivial paths are closed, the
covering is said {\it unbranched}. Otherwise it has branch points: these are
points in $\Sigma_T$ such that a closed loop around them, when lifted, goes
from one sheet of the coverings to a different one. Therefore each branch point
is associated to a non trivial permutation of the $N$ sheets of the coverings.
\par
Consider now a cell decomposition of the target surface $\Sigma_T$ , made of
$\Nc_0$ sites, $\Nc_1$ links and $\Nc_2$ plaquettes, so that the Euler formula
gives the genus $G$ as
\be
\Nc_0-\Nc_1+\Nc_2=2-2G~.
\eeq
To construct a branched $N$-covering of this discretized target surface,
consider $N$ copies of each plaquette $p$ of $\Sigma_T$: these will be the
plaquettes of $\Sigma_W$. Consider now two neighboring plaquettes $p_1$ and
$p_2$ in $\Sigma_T$, let $l$ be their common link, and glue each of the $N$
copies of $p_1$ to one of the $N$ copies of $p_2$ along $l$ according to a
permutation $P\in S_N$. Repeat the procedure for all links to construct the
discretized version of the covering surface $\Sigma_W$.
\par
\begin{figure}
\begin{center}
\epsfxsize 9cm
\epsffile{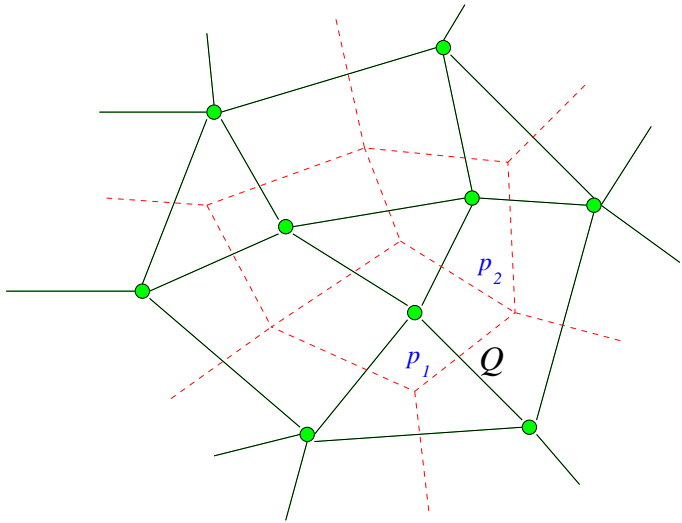}
\end{center}
\caption{Dual lattice: the solid lines are the links of the original
lattice and the dashed lines the links of the dual lattice. The
plaquettes $p_1$ and $p_2$ are joined by a dual link to which a
permutation $Q\in S_N$ is associated: $Q$ dictates how to glue
together copies of $p_1$ to copies of $p_2$ to construct the covering surface.}
\label{Figure:dual}
\end{figure}

In this way one associates a permutation $P\in S_N$ to each link of $\Sigma_T$.
It is easier to visualize $P$ as associated to an oriented link of the {\it
dual} lattice, that is the one in which sites and plaquettes are exchanged, as
shown in Fig.~\ref{Figure:dual}. A dual link goes from a plaquette $p_1$ to a
plaquette $p_2$ and its associated permutation dictates how to glue copies of
$p_1$ to copies of $p_2$.
\par
A closed path around a site $s$ of $\Sigma_T$ will be represented in the
discretized version as a plaquette of the dual lattice. If the ordered product
of the permutations around the plaquette is not the identical permutation, then
$s$ is a branch point of our covering: the liftings of the closed path to
$\Sigma_W$ are not all closed, as some of them start on a sheet and end on a
different one.  Therefore in our discretization the branch points of the
coverings are localized on the lattice sites.
\par
\begin{figure}
\begin{center}
\epsfxsize 6.8cm
\epsffile{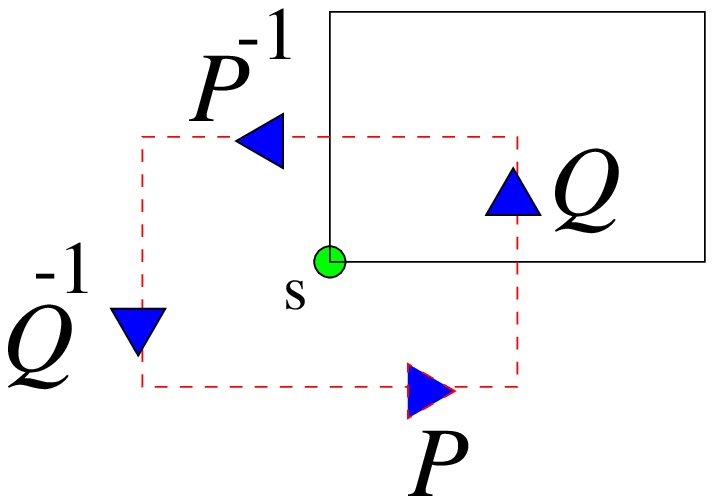}
\null\hskip 0.8cm
\epsfxsize 4.8cm
\epsffile{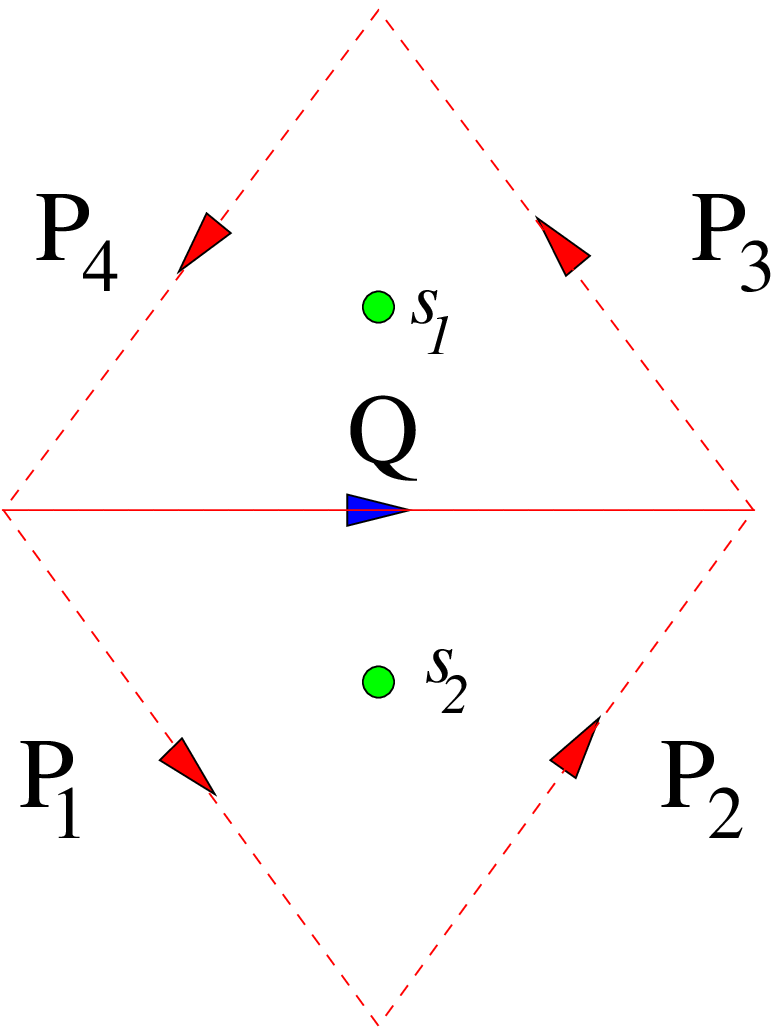}
\end{center}
\caption{\emph{Left:} One-plaquette discretization of the torus. The solid 
lines are the original lattice links exiting from the site $s$. The dashed
lines are the dual links, to which the permutations $P$ and $Q$ are associated.
A closed path around the site $s$ gives rise to the permutation
$PQP^{-1}Q^{-1}$, so that if this is not the identical permutation then $s$ is
a branch point. \emph{Right:} Integration over internal (dual) links.}
\label{Figure:torus}
\end{figure}
Now we want to construct a model in which all possible branched $N$-coverings
of a given (discretized) target surface are counted with Boltzmann weights
depending on their branch point structure. The previous discussion suggests
that this model can be written as a lattice gauge theory with $S_N$ as the
gauge group.
\par
The possible types of branch points are in one-to-one correspondence with the
conjugacy classes of permutations, that is with the partitions of $N$. To each
conjugacy class we want to assign an arbitrary, positive Boltzmann weight.
Therefore we must construct a lattice gauge theory of the symmetric group,
defined on the discretized target surface: the Boltzmann weight of a
configuration  will be
\be
\prod_s w_s(P_s)~,
\label{weight}
\eeq
where the product is extended to all the sites, $P_s\in S_N$ is the ordered
product of the permutations on the dual links around the site $s$, that is
around the dual plaquette corresponding to the site $s$ and the weights
$w_s(P_s)$ vary in general from site to site and depend on the conjugacy class
of $P_s$ only, namely:
\be
w_s(P)= w_s(QPQ^{-1}) \ \ \ \forall\ P,Q\in S_N~.
\label{classfun}
\eeq
\par
The partition function of our model is then
\be
Z_N=\left(\frac{1}{N!}\right)^{\Nc_2}\sum_{\{P\}}\prod_{s}w_s(P_s)~,
\label{zeta}
\eeq
where the sum is extended to all the configurations, that is to all ways of
assigning a permutation to each link, and the product to all sites. The
normalization takes into account the arbitrary relabelings of  the sheets on
top of each plaquette.
\par
This partition function  can be computed by first expanding the class function
$w_s(P)$ in characters of the symmetric group:
 \ba
w_s(P)&=&\frac{1}{N!} \sum_r d_r \tilde{w}_s(r) \ch_r(P)\label{charaction1}\\
\tilde{w}_s(r)&=&\sum_{P\in S_N} \frac{\ch_r(P)}{d_r} w_s(P)~.
\label{wr}
\ea
We then follow the standard method used for solving two-dimensional lattice
gauge theory, first introduced in Ref.~\cite{Migdal:1975zg,Rusakov:1990rs}.
That is, we use the orthogonality and completeness properties of the characters
(already implicitly used in deriving (\ref{charaction1}) and (\ref{wr})), which
in the case of the symmetric group read:
\ba
\sum_r \ch_r(Q)\ch_r(P) &=& \sum_R \delta(PRQ^{-1}R^{-1})~, \label{compl}\\
\frac{1}{N!}\sum_{P\in S_N}\ch_r(P_1P)\ch_{r^\prime}(P^{-1}P_2)&=&
\delta_{rr^\prime}\frac{\ch_r(P_1P_2)}{d_r}~,\label{ortho1}\\
\frac{1}{N!}\sum_{P\in S_N}\ch_r(PP_1P^{-1}P_2)&=&
\frac{\ch_r(P_1)\ch_r(P_2)}{d_r}~.\label{ortho2}
\ea
\par
\vskip0.5cm
These properties can be used to integrate over all the internal dual links of
the discretized surface. Suppose for example we want to integrate over $Q$ as
shown in Fig.~\ref{Figure:torus}(right). Using the character expansion
Eq.~(\ref{charaction1}) and the orthogonality property Eq.~(\ref{ortho1}), we
find 
\be
\sum_{Q}w_{s_2}(P_1P_2Q^{-1})w_{s_1}(QP_3P_4)=
\frac{1}{N!}\sum_r d_r \tilde{w}_{s_1}(r)\tilde{w}_{s_2}(r)
\ch_r(P_1P_2P_3P_4)~.
\eeq
In this way one can integrate over all the internal links, and end up with an
effective one-site model, whose form depends on the topology of the target
surface. Let us consider first the partition function on the disk: if $P$ is
the permutation around the boundary we get
\be
Z_{N,{\rm disk}}(P)=(N!)^{-2}\sum_r
d_r\ch_r(P)\prod_{s=1}^{\Nc_0}\tilde{w}_s(r)~.
\label{disk}
\eeq
From this expression one can calculate the partition function for a surface
without boundaries and arbitrary genus $G$. This is done by representing the
surface as a polygon with sides suitably identified:
\be
Z_{N,G}=\sum_{P_1,Q_1,\dots,P_G,Q_G}Z_{N,{\rm
disk}}\left(P_1Q_1P_1^{-1}Q_1^{-1}\dots
P_GQ_GP_G^{-1}Q_G^{-1}\right)~.
\eeq
At this point one can use Eqs.~(\ref{ortho1},\ref{ortho2}) to perform the sum
over $P_1,Q_1,\dots,P_G,Q_G$ and obtain the final form of the partition
function:
\be
Z_{N,G}=\left(N!\right)^{2G-2}\sum_r d_r^{2-2G} \prod_{s=1}^{\Nc_0}
\tilde{w}_s(r)
\label{final}
\eeq
\par
This expression%
\footnote{This partition function has been studied already in the literature,
for instance in connection with the interpretation of two dimensional
Yang-Mills theories as string theories \cite{Gross:1993hu}}
allows one to calculate the number of coverings with any prescribed branch
point number and structure. As an example, let us consider the limiting case of
unbranched $N$-coverings: this is obtained by choosing
\ba
w_s(P)=\delta(P)\ \ \ \forall s
\ea
In this case eq.~(\ref{wr}) gives simply
\be
\tilde{w}_s(r)= 1
\eeq
so that
\be
Z_{N,G}=\sum_r \left(\frac{d_r}{N!}\right)^{2-2G}
\label{nobranch}
\eeq
and in particular for the torus we obtain the well-known result
\be
\label{zn1}
Z_{N,1}=\sum_r 1=p(N)~,
\eeq
where $p(N)$ is the number of partitions of $N$.
\par
The previous example can be generalized to coverings with any branch point
structure by associating to each site $s$ a permutation $Q_s$ and choosing the
Boltzmann weight $w_s(P)$ of the form:
\be
w_s(P)=\hat{\delta}(P,Q_s)=\frac{1}{N!} \sum_{R \in S_N} \delta(PRQ_sR^{-1})~,
\label{bweight}
\eeq
which means that $w_s(P)$ is different from zero only if $P$ is in the same
conjugacy class as $Q_s$. From (\ref{bweight}) and the orthogonality of the
characters we get:
\be
\tilde{w}_s(r)= \frac{\ch_r(Q_s)}{d_r}~.
\label{wtilde}
\eeq
By inserting (\ref{wtilde}) into (\ref{disk}) and (\ref{final}) we
obtain:
\be
Z_{N,{\rm disk}}(P,\{Q_s\}) = \frac{1}{N!^2} \sum_r d_r
\ch_r(P)\prod_s\left(\frac{\ch_r(Q_s)}{d_r}\right)
\label{zdisk}
\eeq
and
\be
Z_{N,G}(\{Q_s\})= \sum_r\left(\frac{d_r}{N!}\right)^{2-2G}
\prod_s\left(\frac{\ch_r(Q_s)}{d_r}\right)~.
\label{zG}
\eeq
The r.h.s. of both eq.(\ref{zdisk}) and eq. (\ref{zG})  depend only on the
equivalence  classes of $Q_s$ and ultimately on  how many branch point of each
equivalence class are present on the surface. Let us then denote by $\hat{Q}$ 
the equivalence classes, and by $p_{\hat{Q}}$ the number of branch points in
the equivalence class $\hat{Q}$;  then eqs. (\ref{zdisk}) and (\ref{zG}) can be
rewritten as:
\ba
Z_{N,{\rm disk}}(P,p_{\hat{Q}})&=& \frac{1}{N!^2}\sum_r d_r \ch_r(P)
\prod_{\hat{Q} \neq \hat{\bf 1}}
\left(\frac{\ch_r(\hat{Q})}{d_r}\right)^{p_{\hat{Q}}}~;
\label{zdisk2}\\
Z_{N,G}(p_{\hat{Q}})&=&N!^{2G-2} \sum_r d_r^{2-2G-p} \prod_{\hat{Q}\neq 
\hat{\bf 1}}
\left(\ch_r(\hat{Q})\right)^{p_{\hat{Q}}}~, \label{zG2}
\ea
where $p = \sum_{\hat{Q}\neq \hat{\bf 1}}p_{\hat{Q}}$ is the total number of
branch points. In eq.s (\ref{zdisk2}) and (\ref{zG2}) the numbers $p_{\hat{Q}}$
of branched points of a given type are kept fixed. Suppose now to consider the
numbers $p_{\hat{Q}}$ as additional degrees of freedom, and to let them vary.
This leads us to consider a  partition function, that in a broad sense we can
call grand-canonical, given by
\be
Z_{N,{\rm disk}}(P,{\cal A}) = \sum_{\{p_{\hat{Q}}\}} {\cal A}^p
\prod_{\hat{Q}\neq\hat{\bf 1}}
\frac{\left(\sigma_{\hat{Q}} g_{\hat{Q}}\right)^{p_{\hat{Q}}}}{p_{\hat{Q}}!}
Z_{N,{\rm disk}}(P,p_{\hat{Q}})~,
\label{gcdisk}
\eeq
where $\cal A$ is the area of the disk, $\sigma_{\hat{Q}}$ the number of
permutations in the conjugacy class $\hat{Q}$, and $g_{\hat{Q}}$ a Boltzmann
weight, referred to the unit area, attached to a branch point characterized by
a  permutation $Q \in \hat{Q}$. The factor $ p_{\hat{Q}}!$ accounts for the
fact that branch points in the same conjugacy class are indistinguishable. The
sum is over all the $p_{\hat{Q}}$ with $\hat{Q}\neq \hat{\bf 1}$. It can be
done explicitely and gives
\be
Z_{N,{\rm disk}}(P,{\cal A}) =\frac{1}{N!^2}\sum_r d_r \ch_r(P)
e^{{\cal A} g_r}~, 
\label{gcdisk2}
\eeq
with
\be
g_r = \sum_{Q \neq {\bf 1}} g_{Q} \frac{\ch_r(Q)}{d_r}~,
\label{cierre}
\eeq
where of course $g_{Q}=g_{\hat{Q}}$ if $Q \in \hat{Q}$. In eq, (\ref{cierre})
the sum over the conjugacy classes $ \hat{Q}$ has been replaced by the sum over
the group elements, using the fact that according to the definition of
$\sigma_{\hat{Q}}$ we have $\sum_{\hat{Q}} \sigma_{\hat{Q}} \equiv \sum_{Q}$.
The quantity $g_r$ in (\ref{gcdisk2}) can be thought of as an arbitrary
function  of the representation $r$.
\par
The partition function $Z_{N,{\rm disk}}(P,{\cal A})$ can be used as the 
building block of a different lattice gauge theory of the group $S_N$. Consider
once again a cell decomposition of a Riemann surface of genus $G$, and
associate an element of $S_N$ to each link. Let us now associate to each
plaquette $\alpha$ of area ${\cal A}_{\alpha}$ in our cell decomposition a
Boltzmann weight given by $Z_{N,{\rm disk}}(P_{\alpha},{\cal A}_{\alpha})$, 
where $P_{\alpha}$ is the ordered product of the link elements around $\alpha$.
Unlike the original theory defined in (\ref{zeta}), this theory has no branch
points on its sites, but a dense distribution of branch points (characterized
by $g_r$ or equivalently by $g_Q$) on each plaquette. The integration over the
link variables can be performed again by means of the orthogonality properties
of the characters, and the result for a Riemann surface of genus $G$ is
\be
Z_{N,G}({\cal A}) = \sum_r \left(\frac{d_r}{N!}\right)^{2-2G} e^{{\cal A} g_r}~.
\label{zetag}
\eeq
We can regard the theory introduced at the beginning of this section, with the
branch points at the sites of the cell decomposition as a ``microscopic"
theory, and the one defined by the partition function (\ref{zetag}) as a
``continuum" limit. The theory characterized by the plaquette action
(\ref{gcdisk2}) has the same structure as a generalized two dimensional YM
theory, but with $S_N$ as gauge  group. As a special case we can consider the
one where all branch points are quadratic. This is obtained from (\ref{cierre})
by choosing 
\be
g_r=g\, \xi^r_2 \equiv g \frac{N(N-1)}{2}\frac{\ch({\bf 2}^1)}{d_r}~,
\label{quadrbr}
\eeq
where  ${\bf 2}^1$ donotes the conjugacy class consisting of just one exchange.
$\xi^r_2$  can be expressed in terms of the lengths $m_{\alpha}$ and 
$n_{\beta}$ of the rows and the columns of the Young tableau labeling the
representation $r$ as $\xi^r_2=1/2 ( \sum_{\alpha}m_{\alpha}^2 -  \sum_{\beta}
n_{\beta}^2)$. It is related to the quadratic Casimir of the representation of
a unitary group corresponding to the same Young tableau.
\par
\begin{figure}
\begin{center}
\epsfxsize 6.5cm
\epsffile{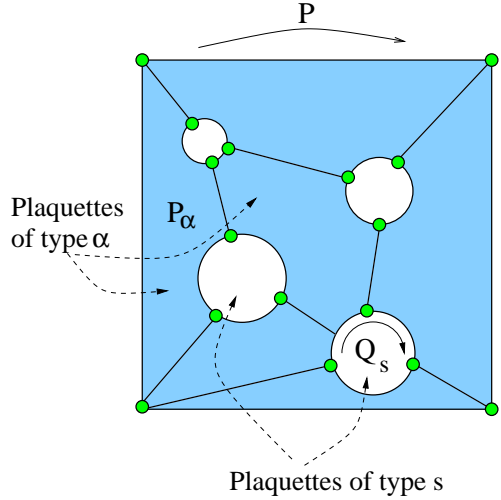}
\end{center}
\caption{Cell decomposition of a disk with $p$ holes ($p=4$ in this case).}
\label{Figure:holes}
\end{figure}

The statistics of branched coverings with a finite number $p$ of branched 
points, as summarized in eq.s (\ref{zdisk2}) and (\ref{zG2}), can  also be
viewed in a slightly different way. Consider a disk with $p$ holes, and let the
holonomies on the internal and external boundaries be respectively $Q_s$ and
$P$. A cell decomposition of this surface is shown in Fig.~\ref{Figure:holes} 
and an element of $S_N$ can be associated to each link  of the cell
decomposition. Assuming that there is no branch point on the surface we
associate to each plaquette $\alpha$ the Boltzmann weight
\be
w_\alpha = \delta(P_{\alpha})~,
\label{plaq}
\eeq
where $P_{\alpha}$ is the ordered product of the elements of $S_N$ associated
to the links around the plaquette $\alpha$. There is a second type of
plaquettes  in the cell decomposition that we are considering, namely the
plaquettes that correspond to the boundaries of the holes. We can choose the
holonomies on these  boundaries to be fixed and given by $Q_s$ as in eq.
(\ref{zdisk}). This is equivalent to associate the Boltzmann weight
$w_s(P_s)=\hat{\delta}(P_s,Q_s)$  introduced in eq.(\ref{bweight}) to the
corresponding plaquettes. We have in this way constructed an $S_N$ lattice
gauge theory with two types of plaquettes, the ones obtained by the cell
decomposition of the surface, and the ones corresponding to the holes, which
are in fact nothing else but blown-up branched points. Clearly, the partition
function of such lattice theory  is again given by (\ref{zdisk2}). Instead, if
we associate to all boundary plaquettes the same  Boltzmann weight
$w_s(P_s)=\sum_r d_r {\cal A} g_r \ch_r(P_s)$ and we  sum over the number  of
boundaries we reproduce  the "continuum" partition function (\ref{zetag}).
\par
This construction is not obviously limited to the gauge group $S_N$. Consider
for instance YM theory on a surface of genus $G$ with $p$ boundaries, with a
cell decomposition anologous to the one of the disk in
Fig.~\ref{Figure:holes}.  We associate to  the internal plaquettes $\alpha$ a
Boltzmann weight
\be
w_\alpha(g_{\alpha}) = \sum_R d_R \chi_R(g_{\alpha})~,
\label{bfpl}
\eeq
where $g_{\alpha}$ is the ordered product of the link variables around the
plaquette. This  defines on the surface  a topological (BF) theory. To each of
the $p$ plaquettes forming the boundaries we associate instead a Boltzmann
weight
\be
w_s(g_s) =   \sum_R d_R C_R {\cal A} \chi_R(g_s)~,
\label{bapl}
\eeq
\section{Matrix strings of generalized YM2, coverings, and the lattice gauge
theory of $\mathcal{G}_N$}
As already discussed in the introduction, the quantization in the unitary gauge
of a generalized U$(N)$ Yang-Mills theory is described
\cite{Billo:1999fb,Billo:2000ts} by a theory of coverings of the space--time
manifold, {\em i.e.} a string theory. This theory is however not the pure
theory of coverings that we described in the previous section as a lattice
gauge theory of the permutation group $S_N$. In fact a U$(1)$ gauge theory is
defined on the world-sheet of the string. This is  the result of the residual
gauge freedom after choosing the unitary gauge in which the auxiliary field $B$
is diagonal.
\par
We have already shown in \cite{Billo:2000ts} that in the case of
\emph{unbranched} coverings a string theory with U$(1)$ gauge fields on the
world sheet can be described as  a lattice gauge theory in which the gauge
group is the semidirect product $\mathcal{G}_N$ of $S_N$ and U$(1)^N$. The
elements of this group, which we denote by $(P,{\bf \phi})$, can be represented
by U$(N)$ matrices of the special form
\beq
\label{sr1}
(P,\phi)_{ij} = \ee^{\ii\phi_i}\, \delta_{i,P(j)}~,
\eeq
where $P\in S_N$.
\par
In order to understand how the lattice gauge theory of the group
$\mathcal{G}_N$ originates, let us go back to Fig. 1, at the beginning of Sec.
2, and consider again $N$ copies of each plaquette $p$. Although we are
interested here in introducing on the world sheet a gauge group U$(1)$, we
consider the case of an arbitrary gauge group $G$ to make the construction 
completely general. Let us introduce on each plaquette $p$ a matter field
$\Psi_i^{\alpha}(p)$ that transform under a given representation of $G$. In
$\Psi_i^{\alpha}(p)$ the index $i = 1,2,\dots,N$ labels the copies of the
plaquette and $\alpha$ is the index of the representation of $G$. A gauge
transformation consists in a relabeling of the sheets induced by a permutation
$P$ and, on each sheet $i$, in the gauge transformation induced by $g_i\in G$,
namely:
\be
\Psi_i^{\alpha}(p) \longrightarrow (P,{\bf g}) \Psi_i^{\alpha}(p)=D^{\alpha}_
{\beta}(g_i)\Psi_{P^{-1}(i)}^{\beta}(p)~,
\label{gaugetr}
\eeq
where $D^{\alpha}_{\beta}(g_i)$ denote the matrix elements of $g_i$ in the
given representation. From eq. (\ref{gaugetr}) the composition rule of two
elements of the gauge group can be easily derived and found to define the
semidirect product of $S_N$ and $G^N$:
\be
(P,{\bf g})(Q,{\bf h})=(PQ,{\bf g}\cdot P{\bf h})~~~~~~~{\rm with}~~~~~~~
({\bf g}\cdot P{\bf h})_i=g_i h_{P^{-1}(i)}~.
\label{grmolt}
\eeq
From (\ref{grmolt}) one also obtains:
\be
(P,{\bf g})^{-1} = (P^{-1},P^{-1}{\bf g}^{-1})~.
\label{invrs}
\eeq
Consider now the dual lattice. The matter fields sit on the sites $s$ of the
dual lattice, while on the links we have to define gauge connections,  given by
group elements, which are needed to define covariant differences, the discrete
analogue of covariant derivatives. Non trivial holonomies are associated to the
plaquettes of the dual lattice, namely to the sites of the original lattice,
and are given by the ordered product along  the plaquette of the group elements
on the links. Let $(P_s,{\bf g}(s))$ be such product relative to a plaquette
$s$. A gauge transformation on the plaquette variable $(P_s,{\bf g}(s))$ is
given by:
\be
(P_s,{\bf g}(s)) \longrightarrow (Q,{\bf h})(P_s,{\bf g}(s))(Q,{\bf h})^{-1}=
(QP_sQ^{-1},{\bf h}\cdot Q{\bf g}\cdot QPQ^{-1}{\bf h^{-1}})~.
\label{gtr}
\eeq
For the $S_N$ part, this gauge transformation amounts, as in the model of the
previous section, to a relabeling of the sheets and leaves unchanged the
decomposition into cycles of $P_s$, which describes the branching structure of
the point $s$. As for the $G$ gauge transformations, they can be derived from
(\ref{gtr}) by setting $Q={\bf 1}$, and read
\be
g_i(s) \longrightarrow h_i \cdot g_i(s) \cdot h_{P^{-1}(i)}^{-1}~.
\label{ggtr}
\eeq
Clearly, unless $P^{-1}(i)=i$, the transformation given in eq.(\ref{ggtr}) is
not a gauge transformation on $g_i(s)$. This is due to the fact that the
lifting of a closed loop around the point $s$ on the target space $\Sigma_T$ is
in general not closed. The closed loops on the world sheet $\Sigma_W$ are given
by the cycles of $P_s$, and correspondingly the gauge covariant loop variables
are the products $g_{i_1}(s)\cdot g_{i_2}(s)\cdot \ldots g_{i_k}(s)$ where
$(i_1,i_2,\ldots ,i_k)$ is a cycle of $P_s$.
\par
We have shown that the theory of branched $N$-coverings with  a $G$-gauge
theory of on the world sheet is equivalent to a lattice gauge theory where the
gauge  group is the semi-direct product of $S_N$ and $G^N$. In order to write
the  partition function of this theory, we need to find the irreducible 
representations of such group.  Although this can be done for an arbitrary
group $G$, we shall restrict ourselves in the rest of this section to the case
$G={\rm U}(1)$. The extension to arbitrary groups, although cumbersome from the
point of view of notations, is rather straightforward.
\subsection{Representation theory of $\mathcal{G}_N$}
Consider now the group $\mathcal{G}_N$, the semidirect product of $S_N$ and
U$(1)^N$, whose elements we denote by $(P,\phi)$ where $P$ is an element of
$S_N$ and $\phi$ stands for the set of invariant angles
$\phi_i~~~~i=1,2,\ldots,N$ that characterize an element of U$(1)^N$. The
product of two generic elements of the group can be obtained from (\ref{sr1})
or directly from (\ref{grmolt}) by putting $g_i=\ee^{\ii \phi_i}$:
\beq
\label{sr2}
(P,\phi)(Q,\theta) = (PQ, \phi + P\theta)~,
\eeq
where $(P\theta)_i = \theta_{P^{-1}(i)}$. With this product law, the inverse of
an element is
\beq
\label{sr3}
(P,\phi)^{-1} = (P^{-1},-P^{-1}\phi)
\eeq
and the expression of a conjugated element is
\beq
\label{extrasr1}
(Q,\theta)(P,\phi)(Q,\theta)^{-1} = (QPQ^{-1},\theta+Q\phi - QPQ^{-1} \theta)~.
\eeq
\par
The structure of the group $\mathcal{G}_N$ is similar to that of the
Poincar\'e group, the $U(1)^N$ elements playing the role of the translations
and the permutations the role of the Lorentz rotations. To describe the
irreducible representations of $\mathcal{G}_N$ we can thus follow Wigner's
method of induced representations usually employed for the Poincar\'e group.
\par
We begin by choosing an irreducible representation  $\n \equiv\{n^i\}$ of
U$(1)^N$, with $i=1,\ldots,N$ and $n^i\in \mathbb{Z}$. This is nothing but the
assignment of quantized momenta in all of the $N$  compact directions: the
U$(1)^N$ element $\phi$ is represented by the  phase $\exp(\ii\sum_i n^i
\phi_i)$. The chosen \emph{unordered} $N$-ple of momenta is invariant under the
action of the permutation group; however, this is not the case for the specific
ordering of them which defines the representation $\n$. We use again the 
notation
\beq
\label{sr4}
P\n = \{n^{P^{-1}(i)}\}
\eeq
to denote a permutation of the momenta. In the analogy with the Poincar\'e
group, the \emph{unordered} $N$-ple of the  $n^i$'s is the invariant that plays
the role of the squared mass.
\par
Given the representation $\n$, Wigner's idea is to construct the representations
of the semidirect product group in terms of irreducible representations of the
so-called \emph{little group} $\mathcal{L}_\n\subset S_N$,  which  contains
those permutations that preserve $\n$. We can split any permutation $P\in S_N$
as
\begin{equation}
\label{semi5}
P=\pi\,p\,:~~\left\{\begin{array}{l}
p\in\mathcal{L}_n~, \\
\pi\in S_N/\mathcal{L}_\n~.\end{array}\right.
\end{equation}
While $p\n=\n$, the elements $\pi$ in the coset  $S_N/\mathcal{L}_\n$ act
non-trivially, mapping $\n$ into $\pi \n$, with
\begin{equation}
\label{semi6}
(\pi \n)^i = n^{\pi^{-1}(i)}~,
\end{equation}
in accordance with \eq{sr4}. Notice that the coset class $\pi$ is of course
defined only up to  little group transformations. We can single out a specific
representative $\hat\pi$ in this class by requiring, for instance, that
\beq
\label{repchoice}
\mbox{$\hat\pi(i)<\hat\pi(j)$ if $i<j$ with $n^i=n^j$.}
\eeq
This removes any ambiguity from the decomposition \eq{semi5}; where not
otherwise indicated, we assume such a choice in what follows.
\par
The little group is determined, up to isomorphisms, only by the
\emph{structure} of the set $\{n^i\}$ defining the representation. The
construction is the following: let us denote by $N_n$ with $\sum_a N_a = N$ the
number of times a given momentum $n$ appears in the set $\{n_i\}$. In other
words there are $N_n$ values of $i$ for which $n_i=n$. Then the little group
consists of  the direct product
\begin{equation}
\label{extra1}
\bigotimes_n S_{N_n}
\end{equation}
of the symmetric groups $S_{N_n}$ acting on the subsets of indices $i$ for
which $n_i=n$. In the Poincar\'e group there are only two different little
group structures corresponding to the mass squared being bigger than or equal
to zero. In our case we have many more possibilities, specified by the possible
degeneracies amongst the set of $n_i$'s.
\par
The explicit expression $\mathcal{L}_\n$ in \eq{semi5} of the little group
\eq{extra1} as a subgroup of $S_N$ depends on the specific ordering of the
$n_i$'s that defines the representation $\n$. If we modify the ordering $\n$ to
$\pi \n$, we have an isomorphic expression
\begin{equation}
\label{semi7}
\mathcal{L}_{\pi \n}=\pi\,\mathcal{L}_\n\,\pi^{-1}~.
\end{equation}
\par
As it follows from \eq{extra1}, an irreducible representation $\r$ of the
little  group $\mathcal{L}_\n$  is a tensor product of irreducible
representations $r_a$ of the symmetric groups $S_{N_a}$, for $a=1,\ldots M$.
The irreducible representations $r_a$, which are in one--to--one correspondence
with the Young tableux of $N_a$ boxes, and their characters $\ch_{r_a}$ have
been discussed in the previous Section.  We denote the matrix elements of $p\in
\mathcal{L}_\n$ in the representation $\r$ by
\beq
\label{sr5}
[D_r(p)]_\alpha^{~\beta} = \bigotimes_{a=1}^M
[D_{r_a}(p_a)]_{\alpha_a}^{~\beta_a}~,
\eeq
where $p_a$ is the component of $p$ in the $a$-th factor, $S_{N_a}$, of the
little group. Correspondingly, the characters $\Ch_\r(p)$ in this
representation are given by
\beq
\label{sr6}
\Ch_\r(p)= \prod_{a=1}^M \ch_{r_a}(p_a)~.
\eeq
\par
Consider now the action of a permutation $P$ on a state with momenta
$\n$ and which transforms in the irreducible representation $\r$ of the little
group. Denoting such a state as $\ket\alpha^{(\n)}$, we have
\beq
\label{sr7}
P\ket\alpha^{(\n)} = [D_\r(p)]_\alpha^{~\beta} \ket\beta^{(\pi \n)}~,
\eeq
having used the decomposition \eq{semi5} with a specific choice of coset
representative, \emph{e.g.}, the one in \eq{repchoice}.  We see that the states
at fixed $\n$ do not span a representation by themselves, and we are forced to
form a single representation including all of the possible reorderings of the
$n_i$'s, parametrized by the classes of $S_N/\mathcal{L}_\n$. This is analogous
to the obvious fact that a Lorentz rotation $\Lambda$ on a state of momentum
$p^\mu$  produces a state of rotated momentum $\Lambda^\mu_{~\nu} p^\nu$.
\par
\eq{sr7} suggests that we can represent, for each fixed U$(1)^N$ representation
$\n$,  the entire group $\mathcal{G}_N$ on a finite-dimensional space spanned
by state vectors $\ket{\sigma ; \alpha}$, with  $\sigma\in S_N/\mathcal{L}_\n$
and $\alpha$ in the carrier space of an irreducible representation $r$ of the
little group $\mathcal{L}_n$, as above. The matrix representing an element
$(P,\phi)\in \mathcal{G}_N$ can then be written, using the composite index $A =
(\sigma;\alpha)$, as
\begin{equation}
\label{semi9}
[D_{\n,\r}(P,\phi)]_A^{~B} = \ee^{\ii\sum_i (\sigma n)^i \phi_i}\,
[D_\n(P)]_\sigma^{~\tau}~
[D_\r({\hat\sigma}^{-1}P\hat\tau)]_\alpha^{~\beta}~,
\end{equation}
where
\begin{equation}
\label{semi10}
[D_\n(P)]_\sigma^{~\tau} = \delta(P^{-1}\sigma \n,\tau \n)~.
\end{equation}
The $[D_\n(P)]_\sigma^{~\tau}$ factor simply states that we have non-zero
matrix elements in the space of $\sigma,\tau$ indices, i.e. in
$S_N/\mathcal{L}_\n$, $\sigma \n$ (momenta of the state  we are acting on) to
$\tau \n$. This being the case, we see that $\sigma^{-1}P\tau \n = \n$, i.e., 
for any choice of representatives of the  coset classes,  $\sigma^{-1}P\tau$ is
an element of the  little group and can  rightly  appear in the last factor in
\eq{semi9}.  However, which specific element ${\hat\sigma}^{-1}P\hat\tau$ one
gets depends on the choice of representatives, so to fully describe the
representation  matrices we have to make a choice, \emph{e.g.},  the one of
\eq{repchoice}, as indicated in \eq{semi9}. Different choices of coset
representatives lead to equivalent representations.
\par
It is easy to verify that the matrices $D_{\n,\r}$ defined as in \eq{semi9}
provide indeed a representation of the group $\mathcal{G}_N$, i.e., that
\beq
\label{sr8}
[D_{\n,\r}(P,\phi)]_A^{~B}\,[D_{\n,\r}(Q,\theta)]_B^{~C} =
[D_{\n,\r}(PQ,\phi + P\theta)]_A^{~C}~.
\eeq
\par
One can furthermore show that the representations $D_{\n,\r}$ are
\emph{irreducible}, by employing Schur's lemma: if one assumes that a matrix
$F_A^{~B}$ commutes with all the elements $D_{\n,\r}(P,\phi)$, one finds that
$F$ has to be proportional to the identity matrix.
\par
The representations $D_{\mu \n,\r}$, with $\mu\in S_N/\mathcal{L}_\n$, obtained
by  considering a different reordering of the $n^i$'s, can be shown to be
equivalent to the representations $D_{\n,\r}$.
\par
Summarizing, the set of inequivalent irreducible representations of the group
$\mathcal{G}_N$ is described by the possible unordered $N$-ples of momenta
$n^i$  and the irreducible representations of the corresponding little group
$\mathcal{L}_\n$.
\par
The characters in these representations we denote by $\CH_{\n,\r}$ and are
given, as it follows from \eq{semi9}, by
\beq
\label{sr9}
\CH_{\n,\r}(P,\phi) = \sum_A [D_{\n,\r}(P,\phi)]_A^{~A} =
\sum_\sigma\, \ee^{\ii\sum_i (\sigma \n)^i \phi_i}\,
\delta(P^{-1}\sigma \n,\sigma \n) \, \Ch_\r(\sigma^{-1}P\sigma)~.
\eeq
\par
The dimension of the representation $D_{\n,\r}$, which we denote by
$d_{\n,\r}$, is given by $\CH_{\n,\r}(\mathbf{1},0)$, \emph{i.e.}, by
\beq
\label{dim}
d_{\n,\r} = \sum_\sigma\, d_\r
=\left| S_N/\mathcal{L}_\n\right|\, d_\r
=\frac{N!}{\prod_n N_n !}\, d_\r~,
\eeq
where $d_\r$ is the dimension of the chosen irreducible representation $\r$
of the little group.
\par
An extreme case is the one in which all the $n^i$'s are different; in this case
the little group is  trivial and has only the trivial representation ($\r=0)$;
all representations $D_{\n,\r=0}$ have then dimension $N!$, and are
given by:
\beq
[D_{\n,\r=0}(P,\phi)]_Q^R = \ee^{\ii \sum_i (Qn)^i \phi_i}
\delta(P^{-1}Q,R)~,
\label{extreme}
\eeq
where $Q$ and $R$ are indices of $S_N/\mathcal{L}_\n$ which coincides
in this case with $S_N$%
\footnote{Notice that if one omits in
(\ref{extreme}) the U$(1)^N$ phase factors, one obtains the regular
representation of $S_N$ which, however, is reducible.}.
At the other end we have the case where all the $n^i$'s are equal, the little
group coincides with $S_N$, and the only coset class is the identity one. All
representations $D_{\n,\r}$ have the same dimensionality $d_\r$ as the chosen
representation of the little group, and coincide, up to the U$(1)^N$  phase
factor, with the representations of $S_N$.
\par
The characters given in \eq{sr9} satisfy the usual orthogonality and
completeness relations, which we will need to construct and solve the lattice
theory. One has the ``fusion'' rule
\bea
\label{fusion}
& & \int D\phi\, \frac{1}{N!} \sum_{P\in S_N}
\CH_{\n,\r}((Q_1,\theta_1)(P,\phi)^{-1})
\CH_{\n',\r'}((P,\phi)(Q_2,\theta_2))
\nonumber\\
& & =
\delta_{\n,\n'}\,\delta_{\r,\r'}\,
\frac{\CH_{\n,\r}((Q_1,\theta_1)(Q_2,\theta_2))}{d_{\n,\r}}~,
\eea
which in the particular case $Q_1=Q_2={\bf 1}$ and $\theta_1=\theta_2=0$
becomes the orthogonality relation
\beq
\label{orthogn}
\int D\phi \frac{1}{N!} \sum_{P\in S_N} \CH_{\n,\r}((P,\phi)^{-1})
\CH_{\n',\r'}(P,\phi) =\delta_{\n,\n'}\,\delta_{\r,\r'}~.
\eeq
In both equations the measure on U$(1)^N$ is defined as $D\phi \equiv \prod_i
d\phi_i/2\pi$. Besides we have the  following ``fission`` property:
\bea
\label{fission}
& & \int D\phi\, \frac{1}{N!} \sum_{P\in S_N}
\CH_{\n,\r}((P,\phi)(Q_1,\theta_1)(P,\phi)^{-1}(Q_2,\theta_2))
\nonumber\\
& & = \frac{1}{d_ {\n,\r}}\,
\CH_{\n,\r}(Q_1,\theta_1) \CH_{\n,\r}(Q_2,\theta_2)
\eea
\par
and the completeness relation
\beq
\label{completeness}
\frac{1}{N!}\sum_\n \sum_\r \CH_{\n,\r}(Q,\theta)\,\CH_{\n,\r}(P,\phi)=
\delta_{\mathcal{G}_N}[(P,\phi),(Q,\theta)]~. 
\eeq
The $\mathcal{G}_N$-invariant delta-function appearing in the l.h.s above 
is explicitly given by
\bea
\label{gninvdelta}
& &\delta_{\mathcal{G}_N}[(P,\phi),(Q,\theta)] =
\frac{1}{N!}\sum_{R\in S_N} \int D\psi\, \delta((P,\phi)(R,\psi)(Q,\theta)
(R,\psi)^{-1})\nonumber \\
& & ~ = \sum_R \delta(PRQR^{-1})\,\frac{1}{(2\pi)^N} \prod_{l=1}^N
\prod_{A=1}^{r_l} 2\pi \delta(\sum_{\alpha=0}^{l-1} \phi_{l,A,\alpha} +
\sum_{\alpha=0}^{l-1} (R\theta)_{l,A,\alpha})~,
\eea
where we decomposed $P$ into $r_l$ cycles of length $l$ ($l=1,\ldots,N$) and
replaced the index $i=1,\ldots, N$ with the multiindex
\beq
\label{multiindex}
(l,A,\alpha)~,\hskip 0.5cm l=1,\ldots, N~,~~~A=1,\ldots, r_l~,~~~
\alpha=0,\ldots, l-1~.
\eeq
Notice that the $\mathcal{G}_N$-invariant delta-function depends only on the 
sums of the angles $\phi$ belonging to the same cycles of $P$ and on the sums
of the angles $\theta$  belonging to the same cycles of  $R^{-1}PR$, that is of
$Q$. A particular case of (\ref{completeness}) is obtained by putting
$Q={\mathbf 1}$  and $\theta_i=0$, and reads:
\beq
\label{completeness2}
\frac{1}{N!}
\sum_\n \sum_\r d_{\n,\r}\,\CH_{\n,\r}(P,\phi)=\delta(P)\, (2\pi)^N
\delta(\phi_1)\dots \delta(\phi_N)~.
\eeq
All these formulas can be checked by direct although somewhat cumbersome
calculations.
\subsection{Branched coverings endowed with a U$(1)^N$ flux and the lattice
gauge theory of $\mathcal{G}_N$}
We are now ready to generalize the results of \secn{sec:dualcov}, and study a
theory of  branched $N$-coverings of a Riemann surface $\Sigma_T$, endowed with
a U$(1)$ gauge theory on the world sheet $\Sigma_W$. According to the
discussion at the beginning of the present section this will be described by a
lattice gauge theory of $\mathcal{G}_N$, namely by a theory where an element
$(P,\phi) \in \mathcal{G}_N$ is associated to each link in the dual lattice. A
specific  covering is determined now by its branch-point structure \emph{and}
by the values of the U$(1)^N$ fluxes. To each plaquette $s$ of the dual lattice
(that is, to each site of the original lattice) we associate a weight
$W_s(P_s,\phi_s)$, where $(P_s,\phi_s)$ is the ordered product of the elements
of $\mathcal{G}_N$ associated to the links around the plaquette $s$. Such
weights must be class functions, that is they depend on the conjugacy class of
$(P_s,\phi_s)$ only. The partition function is obtained summing over all
configurations:
\be
\mathcal{Z}_{G,N}=\left(\frac{1}{N!}\right)^{\Nc_2}\sum_{\{(P,\phi)\}}
\prod_{s}W_s(P_s,\phi_s)
\label{zetagn}
\eeq
(the normalization is as in \eq{zeta}).
We can expand the weight $W_s(P,\phi)$ into characters of $\mathcal{G}_N$,
\beq
\label{chargn exp}
W_S(P,\phi) = \frac{1}{N!}\sum_{\n,\r} d_{\n,\r} \widetilde{W}_s(\n,\r)
\CH_{\n,\r}(P,\phi)~,
\eeq
where, according to \eq{orthogn}, we have
\beq
\label{Wtilde}
\widetilde{W}_s(\n,\r) = \sum_P \int D\phi\,
\frac{\CH_{\n,\r}((P,\phi)^{-1})}{d_{\n,\r}}
W_s(P,\phi)~.
\eeq
\par
Using the character expansion, it is possible to integrate over all internal
links of the dual lattice to obtain finally, just as in \secn{sec:dualcov},
the partition function on a disk:
\be
\mathcal{Z}_{\mathrm{disk},N}(P,\phi)=(N!)^{-2}\sum_{\n,\r} d_{\n,\r}
\prod_{s=1}^{\Nc_0}\widetilde{W}_s(\n,\r)\CH_{\n,\r}(P,\phi)~,
\label{diskgn}
\eeq
where $(P,\phi)$ is the group element associated to the boundary of the disk,
and from this the partition function for a closed surface of genus $G$:
\be
\mathcal{Z}_{G,N}= \sum_{\n,\r} \left(\frac{d_{\n,\r}}{N!}\right)^{2-2G}
\prod_{s=1}^{\Nc_0}\widetilde{W}_s(\n,\r)~.
\label{finalgn}
\eeq
\par
In \cite{Billo:2000ts} we considered the case of unbranched coverings,
and we showed that it corresponds to the following choice of weights:
\beq
\label{unbrwgn}
\forall s :\hskip 0.5cm W_s(P,\phi) = \delta(P) \sum_{n_i}
\ee^{\sum_{i=1}^N (\ii n_i\phi^i - \mathcal{A}_s \sum_i v(n_i))}~,
\eeq
where $\mathcal{A}_s$ is the area of the dual plaquette $s$.  The lattice gauge
theory corresponding to this plaquette action was shown to be equivalent to the
generalized U$(N)$ YM theory based on a potential $V(F)$ if the finite
potential $\sum_i v(n_i)$ is obtained from $V(\{n_i\})$  by subtracting the
logarithmic divergences that arise when the genus $G$ of the target manifold is
different from 1. The $n_i$'s correspond to the quantized eigenvalues of the
diagonal components of $F$.  With this choice, we get from \eq{Wtilde}
\beq
\label{unbrwtgn}
\widetilde W_s(\n,\r) = \ee^{-\mathcal{A}_s \sum_i v(n_i)}~,
\eeq
that is, the $\widetilde W$'s have no dependence on the little group
representation $r$. The partition function for a closed surface is then simply
\be
\mathcal{Z}_{G,N}= \sum_{\n,\r} \left(\frac{d_{\n,\r}}{N!}\right)^{2-2G}
\ee^{-\mathcal{A} \sum_i v(n_i)}~,
\label{Gsimplegn}
\eeq
where $\mathcal{A}=\sum_s \mathcal{A}_s$ is the total area of the target
surface. In \cite{Billo:2000ts}, the grand-canonical partition function 
$\mathcal{Z}_G(q) \equiv \sum_N \mathcal{Z}_{G,N} q^N$ was investigated by
directly enumerating the coverings and associating to each connected  component
a generalized U$(1)$ partition function 
$\sum_{n\in\mathbb{Z}}\ee^{-\mathcal{A}v(n)}$. A closed expression was given in
the case of the torus, and a list  of the first few terms (lowest values of
$N$) in its $q$ expansion was given in the appendix of \cite{Billo:2000ts} also
for other values of $G$. A closed expression for the grand canonical partition
function $\mathcal{Z}_G(q)$, generalizing the one given in \cite{Billo:2000ts}
for the torus, can be obtained from (\ref{Gsimplegn}) by using eq. (\ref{dim}) 
and the relations $\sum_i v(n_i)=\sum_n N_n v(n)$ and $N=\sum_n N_n$.  We find
that the infinite sums factorize in a product over $n$, namely:
\beq
\mathcal{Z}_G(q) \equiv \sum_N \mathcal{Z}_{G,N} q^N =\prod_n 
Z_G(\ee^{-\mathcal{A}v(n)}q)~,
\label{grcan}
\eeq
where $Z_G(q)$ in the grand-canonical partition function for the unbranched
coverings, that is:
\beq
Z_G(q)=\sum_N Z_{N,G} q^N~,
\label{grcov}
\eeq
and $Z_{N,G}$ is given in (\ref{nobranch}). In Appendix \ref{appendix:unbr},
the torus grand-canonical function $\mathcal{Z}_G(q)$ will be described in more
detail to appreciate the relation between the treatment of
\cite{Billo:2000ts}   and the more general one given here.
\par
Let us go back to the general case of an arbitrary branching structure. We
follow the same pattern as in \secn{sec:dualcov} and associate to each  dual
plaquette $s$ an element $(Q_s,\theta_s)$of $\mathcal{G}_N$. This is done by
assigning to it the weight
\beq
\label{brwgn}
W_s(P,\phi) =
\delta_{\mathcal{G}_N}((Q_s,\theta_s),(P,\phi))
\eeq
where the $\mathcal{G}_N$-invariant delta function is the one defined in
(\ref{gninvdelta}). The permutation $Q_s$ gives the branching structure at the
site $s$ of the original lattice, while the invariant angles $\theta_{s,i}$
determine the U$(1)$ holonomies. Notice however that closed loops on the world
sheet around the site $s$ are in one to one correspondence to the cycles of the
permutation $Q_s$. The corresponding U$(1)$ gauge invariant holonomies are
given by the angles $\theta_{s,\{l,A\}} =\sum_{i \in \{l,A\}} \theta_{s,i}$
where $\{l,A\}$ denotes the $A$-th cycle of length $l$ in $Q_s$. These are the
only angles that appear, according to (\ref{gninvdelta}), in the definition of
the covariant delta function and are the only angles which are left invariant,
according to the general discussion at the beginning of the present section,
under conjugacy transformations of $\mathcal{G}_N$.
From Eq.~(\ref{brwgn}) and the orthogonality of characters we obtain
\beq
\label{brwtgn}
\widetilde W_s(\n,\r)  =
\frac{\CH_{\n,\r}(Q_s,\theta_s)}{d_{\n,\r}}~.
\eeq
The partition function on a disk and on a closed surface of genus $G$ can be
obtained by inserting (\ref{brwtgn}) in (\ref{diskgn}) and
(\ref{finalgn})leading to
\beq
\label{partdiskbr}
\mathcal{Z}_{\mathrm{disk},N}((P,\phi),\{(Q_s,\theta_s)\}) =  (N!)^{-2}
\sum_{\n,\r} d_{\n,\r} \prod_{s=1}^{\Nc_0}\left(\frac{\CH_{\n,\r}(Q_s,\theta_s)}
{d_{\n,\r}} \right)\,\CH_{\n,\r}(P,\phi)
\eeq
and
\beq
\label{partbr}
\mathcal{Z}_{G,N}(\{(Q_s,\theta_s)\}) =(N!)^{2G-2}\sum_{\n,\r} d_{\n,\r}^{2-2G}
\prod_{s=1}^{\Nc_0}\left(\frac{\CH_{\n,\r}(Q_s,\theta_s)}{d_{\n,\r}}\right)~.
\eeq
The continuum limit can be taken on Eq.s~(\ref{partdiskbr}) and (\ref{partbr})
following the same track as in \secn{sec:dualcov}.  Let us assume that each
branch point with holonomy $(Q_s,\theta_s)$ appears with a coupling
$\mathcal{A} g[(Q_s,\theta_s)]$, where  $\mathcal{A}$ is the area of the
surface and $g[(Q_s,\theta_s)]$ is a class function of $\mathcal{G}_N$. Let us
then define $g_{\n,\r}$ as the coefficients of the expansion of 
$g[(Q_s,\theta_s)]$ in characters og $\mathcal{G}_N$:
\beq
g[(Q_s,\theta_s)]=\frac{1}{N!} \sum_{\n,\r} d_{\n,\r} g_{\n,\r} 
\CH_{\n,\r}(Q_s,\theta_s)~.
\label{giesse}
\eeq
If we multiply the partition functions  in (\ref{partdiskbr}) and (\ref{partbr})
by $\prod_s \mathcal{A} g[(Q_s,\theta_s)]$, integrate over $(Q_s,\theta_s)$ and
sum over the number ${\Nc_0}$ of branch points (with a $1/{\Nc_0}!$ factor) we
obtain
\beq
\mathcal{Z}_{\mathrm{disk},N}((P,\phi),\mathcal{A}) = (N!)^{-2} 
\sum_{\n,\r} d_{\n,\r} \CH_{\n,\r}(P,\phi) \ee^{\mathcal{A} g_{\n,\r} }
\label{contdisk}
\eeq
and
\beq
\mathcal{Z}_{G,N}(\mathcal{A})=(N!)^{2G-2}\sum_{\n,\r} d_{\n,\r}^{2-2G}
\ee^{\mathcal{A} g_{\n,\r} }~.
\label{contgenus}
\eeq
The partition function given in (\ref{contgenus}) is very general, as it
corresponds to arbitrary weights for the different types of branch points, and
to arbitrary U$(1)$ holonomies associated to each type of branch point.
Consider now a more specific case, where in all sites $s$ there is either no
branch point or one corresponding to a single exchange. This means that the
coupling $g[(Q,\theta)]$ consists of two terms:
\beq
g[(Q,\theta)] =  g_0[(Q,\theta)] +\lambda g_1[(Q,\theta)]~,
\eeq
where $\lambda$ is a free parameter. The functions $g_0[(Q,\theta)]$ and 
$g_1[(Q,\theta)]$ are the most general class functions of  $\mathcal{G}_N$
with  support, respectively, in $Q={\bf 1}$ and in $Q$ consisting of a single
exchange. Their explicit expression is:
\bea
g_0[(Q,\theta)]&=& \delta (Q) \sum_{\{n_i\}} \ee^{\sum_i \ii n_i \theta_i}
v(n_1,n_2,...n_N) \label{gizero}~, \\
g_1 [(Q,\theta)] &=&\sum_R \delta(RP_{12}R^{-1}Q) \sum_{\{n_i\}}\delta_{n_1,n_2}
f(n_1;n_3,...,n_N) \ee^{\ii \sum_i n_i \theta_{R(i)}}~, \label{giuno}
\eea
where $P_{12}$ is a permutation consisting of the exchange of the labels $1$
and $2$, $v(n_1,n_2,...n_N)$ and $f(n_1;n_3,...,n_N)$ are arbitrary functions,
which are symmetric respectively under permutations of all the $n_i$'s and of
the  $n_i$'s with $i=3,4,...N$.  We consider now the case where the
$f(n_1;n_3,...,n_N)=1$, and  $v(n_1,n_2,...n_N)=\sum_i v(n_i)$.  The former
condition implies that there is no U$(1)$ holonomy  attached to the quadratic
branch points or, in other words, that the U$(1)$  electromagnetic field is not
localized on the branch points, but distributed  on the world sheet. The
condition $v(n_1,n_2,...n_N)=\sum_i v(n_i)$ is equivalent to the statement that
the U$(1)$ gauge action is local on the world sheet, namely that what happens
on one sheet has no effect on the others. With this choice the coefficients
$g_{\n,\r}$ can be easily calculated, leading to:
\beq
g_{\n,\r}= \sum_n \left( N_n v(n) + \lambda  \xi^{r_{\{n\}}}_2 \right)~,
\label{gienneerre}
\eeq
where as before $N_n$ is the number of times the integer $n$ appears in $\n$,
$r_{\{n\}}$ is the representation of $S_{N_n}$ in $\r$, and $\xi^r_2$ is defined
in (\ref{quadrbr}).
We can now insert (\ref{gienneerre}) into (\ref{contgenus}) and consider the
grand canonical partition function, defined as in the case without branch 
points (see \eq{grcan}). It is not difficult to verify that with the choice
(\ref{gienneerre}) for $g_{\n,\r}$, the grand canonical partition function
factorizes in an infinite product over $n$, namely
\beq
\mathcal{Z}_G(\mathcal{A},q) \equiv \sum_N \mathcal{Z}_{G,N}(\mathcal{A}) q^N
=\prod_n Z_G\left(\ee^{-\mathcal{A} v(n)} q,\lambda \mathcal{A}\right)~.
\label{grcan2}
\eeq
Here $ Z_G\left( q, \mathcal{A}\right)$ is the grand canonical partition
function for coverings with quadratic branch points, namely:
\beq
Z_G\left( q, \mathcal{A}\right) \equiv \sum_N Z_{G,N}(\mathcal{A}) q^N =
 \sum_N \sum_{r|S_N} \left(\frac{d_r}{N!}\right)^{2-2G} \ee^{\mathcal{A}
 \xi_2^r}~~q^N
\label{grcanbr}
\eeq
where the second sum at the r.h.s. is over the representations $r$ of $S_N$.
\subsection*{Acknowledgments} We thank M. Caselle for many useful discussions.
%
\appendix
\section{Unbranched coverings}
\label{appendix:unbr}
In this appendix, we make more explicit some points of the discussion, given in
the main text after \eq{Gsimplegn}, of the grand-canonical partition functions
describing the particular case of unbranched covers. 
\par
The partition functions for coverings with a U$(1)$ on their world-sheet were
already discussed, in the unbranched case, in \cite{Billo:2000ts}. In that
paper a procedure was given to construct the grand-canonical partition function
$\mathcal{Z}_G(q) \equiv \sum_N \mathcal{Z}_{G,N} q^N$. One starts from the 
grand-canonical partition function $Z_G(q)$ for the pure coverings, and 
considers the corresponding free energy $F_G(q)$. Each connected world-sheet
covering $k$ times the target is then weighed by a factor of $z(k\mathcal{A})$,
where $z(\mathcal{A}) \equiv \sum_{n\in\mathbb{Z}} \exp(-\mathcal{A} v(n))$ is
the generalized U$(1)$ partition function. This  produces the free energy
$\mathcal{F}_G(q)$  for our theory, which can then be exponentiated to obtain
the partition function $\mathcal{Z}_G(q)$.
\par
In the case of the torus, $G=1$, the free energy counting connected coverings
is given by $F_1(q) = \sum_p\sum_{m|p} (1/m) q^p$ (here $m|p$ means ``$m$
divides $p$''), so that an explicit expression for  $\mathcal{Z}_1(q)$ is
\beq
\label{z1q}
\mathcal{Z}_1(q) = 
\exp \Big(\sum_p\sum_{m|p}\frac 1m q^p\,z(p\mathcal{A})\Big)~.
\eeq
The partition functions $\mathcal{Z}_{1,N}$ at fixed $N$ are then
straight-forwardly obtained by expanding \eq{z1q} to the desired order (see the
Appendix of \cite{Billo:2000ts} for the first few values of $N$). For instance,
for $N=2$ and $N=3$ one gets
\bea
\label{z12}
\mathcal{Z}_{1,2} & = & \frac 12 \bigg(z^2(\mathcal{A}) + 3 z(2\mathcal{A}\bigg)~,
\\
\label{z13}
\mathcal{Z}_{1,3} & = & \frac 16 \bigg(z^3(\mathcal{A}) + 9 z(2\mathcal{A}) + 
8 z(3\mathcal{A})
\bigg)~.
\eea
\par
On the other hand, it was noticed in \cite{Billo:1999fb,Billo:2000ts}
that the partition function \eq{z1q} can be re-expressed as an infinite
product:
\beq
\label{z1infprod}
\mathcal{Z}_1(q) = \prod_{n\in\mathbb{Z}} \prod_{k=1}^\infty \frac{1}{1 -
q^k\ee^{-k\mathcal{A}v(n)}} =
\prod_{n\in\mathbb{Z}} \Bigl(\sum_{N=0}^\infty p(s)\, q^s \ee^{-s\mathcal{A}
v(n)}\Bigr)~,
\eeq
$p(s)$ being the number of partitions of $s$. This coincides with
our general formula \eq{grcan}, which for $G=1$ reduces to
\beq
\label{dom1}
\mathcal{Z}_1(q)= \prod_n\left(\sum_s Z_{s,1}\,q^s\,
\ee^{-s\mathcal{A}v(n)}\right)~,
\eeq
since for the pure coverings of the torus one has $Z_{s,1}=p(s)$, as given in
\eq{zn1}.
\par
The equality between the expressions \eq{z1q} and \eq{z1infprod}  of the
grand-canonical partition function summarizes the   rearrangements by which,
for any fixed $N$, one reconstructs from the  sum over the momenta $\n$
appearing in \eq{Gsimplegn} the independent sums over integers which appear in
the expansion of \eq{z1q}. Such an expansion contains in fact products of
U$(1)$ partition  functions $z(k\mathcal{A})$. For instance, let us consider
the case $N=2$. The inequivalent set of momenta $\n$ are: \emph{i)} $n_1 <
n_2$, and \emph{ii)} $n_1 = n_2$. In the case \emph{i)}, the little group is
trivial and so it only has one representation; in case \emph{ii)} the little
group is $S_2$, which admits two irreducible representations. We have therefore
\bea
\label{casen2}
\mathcal{Z}_{1,2} & = & \frac 12 \sum_{n_1\not= n_2} \ee^{-\mathcal{A} (v(n_1) + v(n_2))}
+  2 \sum_{n_1} \ee^{-2\mathcal{A} v(n_1)}
\nonumber\\
& = &  \sum_{n_1,n_2} \ee^{-\mathcal{A} (v(n_1) + v(n_2))} + (2-\frac 12)
\sum_{n_1} \ee^{-2\mathcal{A}v(n_1)} = \frac 12 z^2(\mathcal{A}) + \frac 32
z(2\mathcal{A})~,
\nonumber
\eea
in agreement with \eq{z12}. One can easily repeat the same check for $N=3$ and
(with increasing effort) higher $N$'s.
\par
When the target space has genus $G>1$, in \cite{Billo:2000ts} the expressions
of $Z_{G,N}$ could be worked out case by case, but a closed expression could not
be exhibited. The reason is that no closed form is in fact known for the free
energy $\mathcal{F}_G(q)$ of connected coverings in this case. Here, attacking
the problem by the point of view of the $\mathcal{G}_N$ lattice gauge theory,
we have obtained in \eq{Gsimplegn} such a closed expression. 
\end{document}